\documentclass{statsoc}

\usepackage[a4paper]{geometry}
\usepackage{graphicx}
\usepackage[textwidth=8em,textsize=small]{todonotes}
\usepackage{amsmath}
\usepackage{natbib}
\usepackage{url}
\newcommand{\R}{\textsf{R}~}
\newcommand{\CytoGLMM}{\textsf{CytoGLMM}}
\newcommand{\cytoeffect}{\textsf{cytoeffect}}

\title[Uncertainty Quantification in Multivariate Mixed Models]{Uncertainty Quantification in Multivariate Mixed Models for Mass Cytometry Data}
\author{Christof Seiler}
\address{Department of Statistics, Stanford University; Department of Data Science and Knowledge Engineering, Maastricht University, The Netherlands}
\author[]{Lisa M.~Kronstad, Laura J.~Simpson, Mathieu Le Gars, Elena Vendrame}
\address{Immunology Program and Department of Medicine, Stanford University School of Medicine}
\author[]{Catherine A.~Blish}
\address{Immunology Program and Department of Medicine, Stanford University School of Medicine; Chan Zuckerberg Biohub}
\author[Seiler et al.]{Susan Holmes}
\address{Department of Statistics, Stanford University}

\begin{document}
\begin{abstract}
Mass cytometry technology enables the simultaneous measurement of over 40 proteins on single cells. This has helped immunologists to increase their understanding of heterogeneity, complexity, and lineage relationships of white blood cells. Current statistical methods often collapse the rich single-cell data into summary statistics before proceeding with downstream analysis, discarding the information in these multivariate datasets. In this article, our aim is to exhibit the use of statistical analyses on the raw, uncompressed data thus improving replicability, and exposing multivariate patterns and their associated uncertainty profiles. We show that multivariate generative models are a valid alternative to univariate hypothesis testing. We propose two models: a multivariate Poisson log-normal mixed model and a logistic linear mixed model. We show that these models are complementary and that either model can account for different confounders. We use Hamiltonian Monte Carlo to provide Bayesian uncertainty quantification. Our models applied to a recent pregnancy study successfully reproduce key findings while quantifying increased overall protein-to-protein correlations between first and third trimester.
\end{abstract}

\section{Introduction}

High dimensional flow and mass cytometry enable the simultaneous detection of surface and intracellular proteins at the single cell level. Flow cytometry uses antibodies tagged with fluorochromes and can measure more than 20 proteins on thousands of cells per second \citep{saeys2016computational}. Mass cytometry uses heavy metal isotopes and provides measurements for more than 40 proteins on hundreds of cells per second \citep{bendall2011single}. Immunologists use this new technology to discover new cell types and to study how cells covary across experimental conditions.

There are two main statistical tasks involved in analyzing cytometry data. First, cell types are assembled into latent clusters in a multivariate marker space spanned by Cluster of Differentiation (CD) markers \citep{chan1988simple}; these clusters are labeled according to known cell types. Standard practice is to separate cells sequentially in a hierarchical dichotomy given by positive or negative values along one of the CDs; this is referred to as gating. The sequential approach is helpful in overcoming the curse of dimensionality---the phenomenon where in dimension four or higher the data becomes very sparse and is concentrated at the boundary of the space \citep{orlova2018science,holmes2019modern}. Gating also facilitates biological interpretability. Sequential manual gating is challenging when markers are correlated or higher order interactions of more than three markers are necessary to define a cell type. In such cases, simultaneous clustering for mass cytometry methods \citep{nowicka2017cytof} have gained popularity. Many  clustering methods are readily available \citep{lo2009flowclust,finak2009merging,qian2010elucidation,zare2010data,aghaeepour2011rapid,qiu2011extracting,ge2012flowpeaks,shekhar2014automatic,becher2014high,naim2014swift,meehan2014autogate,van2015flowsom,sorensen2015immunoclust,levine2015data,chen2016cytofkit,samusik2016automated}. \cite{weber2016comparison} provide a benchmark study comparing many of these approaches. To improve interpretability, most researchers use a semi-supervised approach and post-process their clustering results by mapping them to known cell types; e.g.~\citep{spitzer2015interactive}.

The second statistical task is the differential expression analysis between experimental conditions and across cell types. \cite{nowicka2017cytof} propose summarizing protein expressions by taking the median for each cell type. They then use linear models with random effect terms associated both to the donors and the samples and compare the median expression along the different experimental conditions. The statistical power of this approach depends on the number of cell types and samples. There have been efforts to combine both the clustering and differential analysis tasks into one framework. \cite{weber2018diffcyt} extend \cite{nowicka2017cytof} to a larger number of clusters using empirical Bayes moderated tests adapted from transcriptomics. \cite{bruggner2014automated} couple hierarchical clustering and penalized regression to simultaneously find cell types and identify clusters that differentiate between cell types. Their approach divides markers into two types; those that define cell types, and those whose expression levels change in response to stimuli. \cite{arvaniti2017sensitive} abandon the division in cell phenotype and functional markers and use convolutional neural networks to both detect cell types and predict experimental condition, although interpreting the learned network remains a challenge. Finally, \cite{lun2017testing} select differentially abundant clusters between conditions following the spatial false discovery rates literature.

Currently, none of the aforementioned approaches explicitly models marker correlations. Including correlations in the model can have positive effects: it makes the statistical procedure more efficient, it exposes additional structure with which to interpret results, and it provides information as to eventual confounders that need to be attended to.

In this article, we model marker correlations with two mixed models designed specifically for mass cytometry data. We present two complementary approaches. First, a multivariate Poisson log-normal model \citep{aitchison1989multivariate,chib2001markov} that can handle complicated experimental designs, and guards against colliding confounders. Second, a logistic mixed effect model that guards against pipe confounders. Rather than recommending one model, we provide evidence that best results are obtained when we use both models.

In Section~\ref{sec:data}, we describe the publicly available pregnancy data we use throughout this paper. In Section~\ref{sec:models}, we define the two statistical models and introduce the colliding and pipe confounders. In Section~\ref{sec:results}, we show the results of applying our models to the pregnancy data using our two new \R packages \CytoGLMM~and \cytoeffect. In Section~\ref{sec:discussion}, we relate our results to previously reported findings, and comment on possible drawbacks and future challenges.

\section{Data: Pregnancy Study}
\label{sec:data}

We reanalyze a recently published dataset studying the maternal immune system during pregnancy \citep{aghaeepour2017immune}. The study provides a rich mass cytometry dataset analyzed at four time points during pregnancy in two cohorts. The authors isolated cells from blood samples and stimulated them with several activation factors. The goal was to explain how immune cells react to these stimuli, and how these reactions change throughout pregnancy. Findings from such experiments might identify immunological deviations implicated in pregnancy-related pathologies.

The authors collected data at early, mid, late pregnancy, and six weeks postpartum. They kept one sample per donor as an unstimulated control, and stimulated the other samples with a panel of receptor-specific ligands: interferon-$\alpha$2A (IFN$\alpha$), lipopolysaccharide, and a cocktail of interleukins. They processed the samples on a CyTOF 2.0 mass cytometer instrument, and bead normalized the data to account for signal variation over time from changes in instrument performance \citep{finck2013normalization}.

\begin{figure}[t]
\centering
\includegraphics[width=0.65\linewidth]{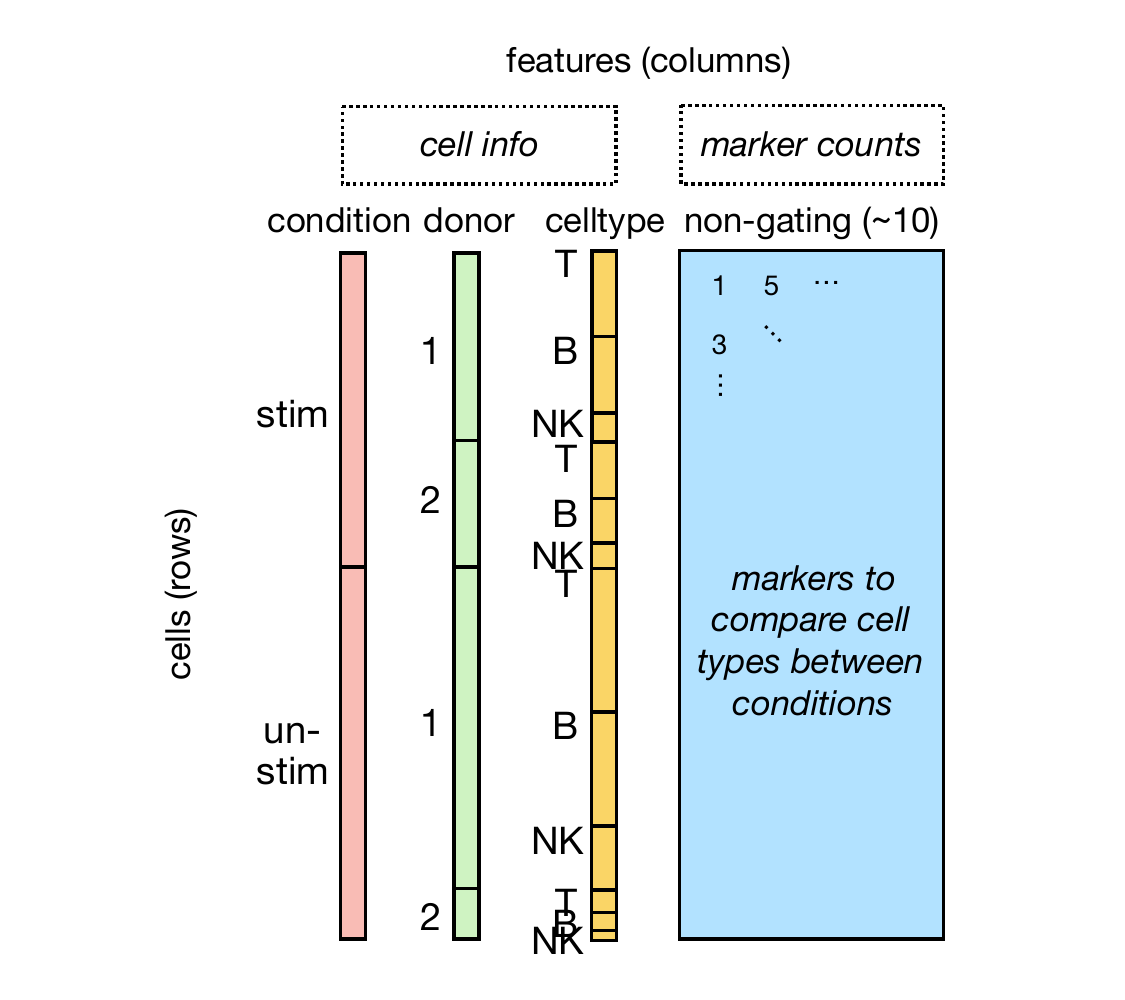}
\caption{Data structure of stimulated and unstimulated cells from two donors after gating to T, B, and NK cell types.}
\label{fig:data_structure_after_gating}
\end{figure}

In our analysis, we focus on comparing early (first trimester) with late (third trimester) pregnancy samples stimulated with IFN$\alpha$ in the first cohort of 18 women. We gate cells into cell types and organize them in a data frame (Figure~\ref{fig:data_structure_after_gating}). We implement this step in our \CytoGLMM~workflow in the supplementary material.

\section{Models}
\label{sec:models}

\subsection{Two Models, Two Interpretations}

In many mass cytometry studies the goal is to explore protein modulation under different stimuli. The underlying mechanisms are often unknown. Models can account for possible confounders and help to guard against spurious correlations. Here we present two regression models: a Poisson Log-normal Mixed Model (PLMM) and a Logistic Linear Mixed Model (LLMM). The PLMM explains marker expressions from experimental conditions. The LLMM does the reverse and ``predicts'' experimental conditions from marker expressions.

\begin{figure}[t]
\centering
\includegraphics[width=0.32\linewidth]{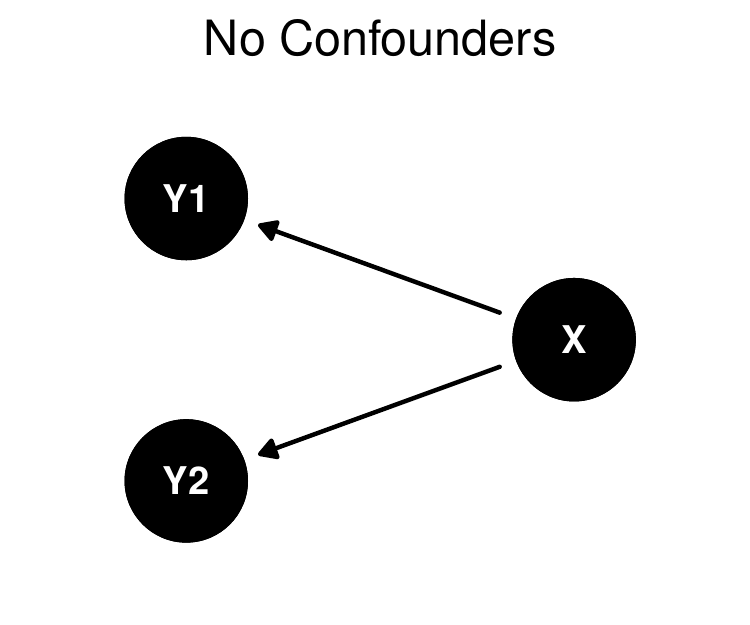}
\includegraphics[width=0.32\linewidth]{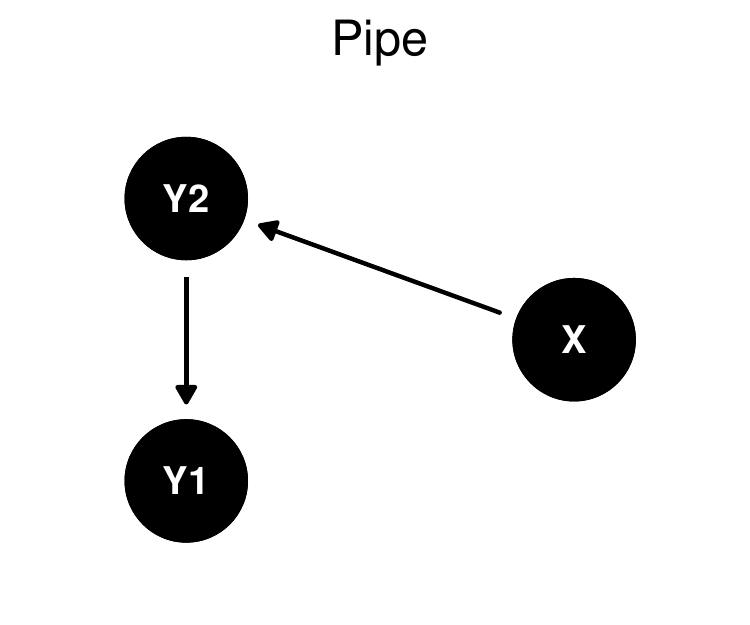}
\includegraphics[width=0.32\linewidth]{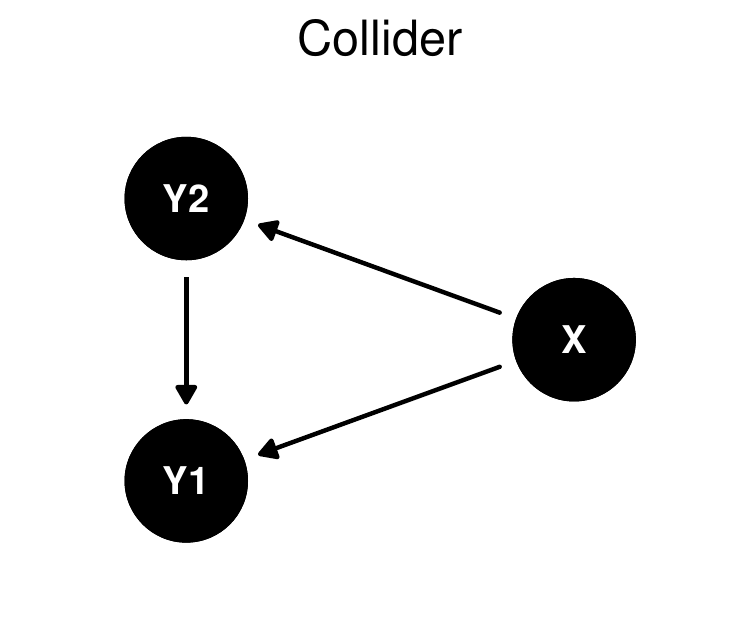}
\caption{Possible causal graphs: no confounders, pipe, and collider. X stands for experimental condition. Y1 and Y2 stand for protein marker expressions.}
\label{fig:dags}
\end{figure}

We use insights from causal inference \citep{spirtes2000causation,rothman2008modern,pearl2009causality,imbens2015causal,peters2017elements,hernan2019causal} to simultaneously interpret results from both models. We distinguish between a pipe and collider confounder. An intuitive way to visualize these concepts is through Directed Acyclic Graphs (DAGs). Assume that X encodes whether a women is in her first or third trimester, and that Y1 and Y2 are two marker expressions. A possible relationship between X and (Y1, Y2) can be a correlation due to a common dependence of expression on the trimester (Figure~\ref{fig:dags}: No Confounders). In such a case both models report similar results. Another possible structure is that trimester causes a change in protein Y2, and Y2 causes a change in Y1 (Figure~\ref{fig:dags}: Pipe). This is sometimes referred to as the pipe confounder, and can be accounted for through a multiple regression model such as our LLMM. In contrast, our PLMM cannot uncover such a relationship and finds that both Y1 and Y2 are correlated with X. In our third construction, we combine both of the previous DAGs. This means that both Y2 and X cause a change in Y1, and X causes a change in Y2 (Figure~\ref{fig:dags}: Collider). This is sometimes referred to as the collider confounder. The danger of colliders is that by conditioning on Y1, as in our LLMM, it can induce spurious negative correlations between X and Y2. In such cases it is more meaningful to use our PLMM \citep{mcelreath2016statistical}.

In our supplementary material we provide simulation results using simplified versions of LLMM and PLMM that demonstrate these effects. Of course, the problem with real data is that we usually have no information about the DAG. In the absence of a known biological mechanism, we therefore recommend fitting both models and giving a combined interpretation of them. In the next two subsection, we present the full models.

\subsection{Poisson Log-Normal Mixed Model}

We now extend the Poisson log-normal mixed model introduced by \cite{chib2001markov}. We begin by assuming that the marker counts $y_{ij}$ in the $i$th cell and $j$th protein marker are distributed according to a Poisson distribution with rate $\mu_{ij}$. Each cell and protein marker has its own rate parameter following a linear model on the $\log$ scale.
The first term $\beta_{\text{cond}[i]j}$ of the linear model are fixed effects in the index notation \citep{mcelreath2016statistical}: $\text{cond}[i] = 1$ and $\text{cond}[i] = 2$ are cells from the first and third trimester, respectively. The second term $b_{\text{cond}[i]ij}$ models a cell-level random effect. This approach can account for correlations between count variables and for cell-to-cell variability. We use the index notation to define multiple cell-level random effects; one per experimental condition. The third term $u_{\text{donor}[i]j}$ models a donor-level random effect. Formally,
\[
\begin{aligned}
y_{ij} & \sim \text{Poisson}(\mu_{ij}) \\
\log(\mu_{ij}) & = \beta_{\text{cond}[i]j} + b_{\text{cond}[i]ij} + u_{\text{donor}[i]j}.
\end{aligned}
\]

Random effect terms $\boldsymbol{b}_{\text{cond}[i]i}$, and $\boldsymbol{u}_{i}$ are vectors of length equal to the number of proteins $J$; in our reanalysis of the pregnancy study $J = 10$. In other studies with less gating markers and more functional markers, these vectors can be $J = 30$ to $J = 40$ dimensional. We have a separate cell-level random effect $\boldsymbol{b}_{\text{cond}[i]i}$ for first $\boldsymbol{b}_{1i}$ and third $\boldsymbol{b}_{2i}$ trimester. We model all three random  effects using independent multivariate normal distributions. Formally,
\[
\begin{aligned}
\boldsymbol{b}_{\text{cond}[i]i} & \sim \text{Normal}(\boldsymbol{0}, \operatorname{diag}(\boldsymbol{\sigma}^{\text{cond}[i]}) \, \boldsymbol{\Omega}^{\text{cond}[i]} \, \operatorname{diag}(\boldsymbol{\sigma}^{\text{cond}[i]})) \\
\boldsymbol{u}_{i} & \sim \text{Normal}(\boldsymbol{0}, \operatorname{diag}(\boldsymbol{\sigma}^{\text{donor}}) \, \boldsymbol{\Omega}^{\text{donor}} \, \operatorname{diag}(\boldsymbol{\sigma}^{\text{donor}})).
\end{aligned}
\]

We choose a weakly informative prior on regression coefficients $\boldsymbol{\beta}$ to rule out highly unlikely cell counts. Within one standard deviation, our prior expects to see a count between $\exp(-7) = 10^{-4}$ and $\exp(7) = 1096$. We might be able to make much stronger assumptions for specific experiments. In most cases, the high number of cells will overwhelm this prior. Formally,
\[
\begin{aligned}
\beta_{\text{cond}[i]j} & \sim \text{Normal}(0,7^2).
\end{aligned}
\]

We choose to use weak priors for the random effects terms. The half-Cauchy for standard deviations $\boldsymbol{\sigma}$ is commonly used \citep{gelman2006prior} as a prior that favors smaller variances and still allows large variances due to its heavy tails. We choose the Lewandowskia, Kurowick, and Joe (LKJ) prior \citep{lewandowski2009generating} with parameter $\eta = 1$ for correlation matrices $\boldsymbol{\Omega}$. This is a uniform prior over correlation matrices. We expect to see quite large correlations and believe that a more flexible prior on the correlation matrices is warranted. In higher dimensions the LKJ will tend to concentrate its mass around zero correlations (see simulations in \cite{seiler2017multivariate}). Formally,
\[
\begin{aligned}
\boldsymbol{\sigma}^{\text{cond}[i]=1}, \boldsymbol{\sigma}^{\text{cond}[i]=2}, \boldsymbol{\sigma}^{\text{donor}} & \sim \text{Half-Cauchy}(0, 2.5) \\
\boldsymbol{\Omega}^{\text{cond}[i]=1}, \boldsymbol{\Omega}^{\text{cond}[i]=2}, \boldsymbol{\Omega}^{\text{donor}} & \sim \text{LKJ}(1).
\end{aligned}
\]

\begin{figure}[t]
\centering
\includegraphics[width=\linewidth]{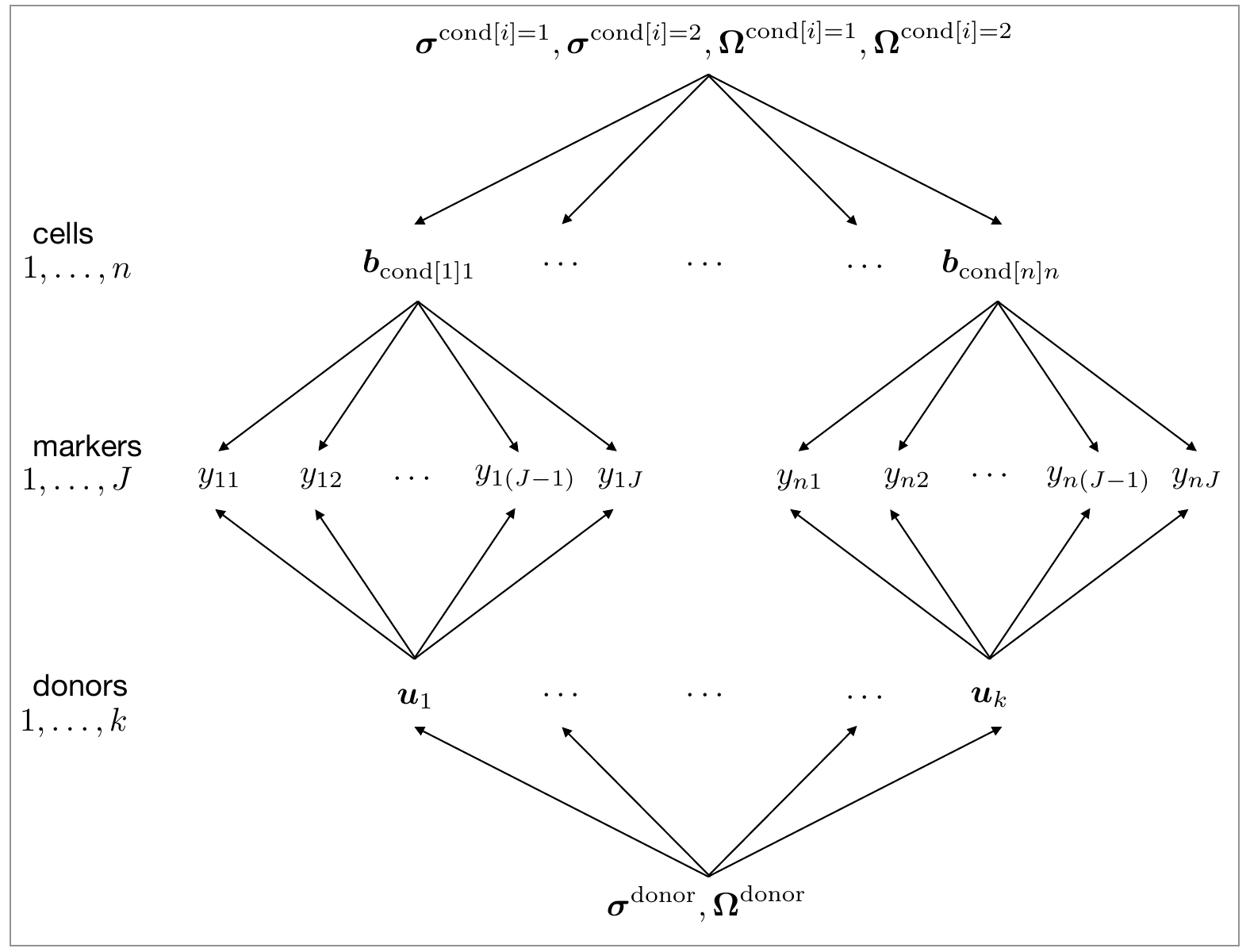}
\caption{A graphical representation of PLMM.}
\label{fig:graphical_plmm}
\end{figure}
Figure \ref{fig:graphical_plmm} displays the PLMM schematically.

To summarize, we extended the Poisson log-normal mixed model by adding separate random effect terms for each experimental condition capturing differences in correlation structure between conditions. We also add a donor random effect term to handle the large donor-specific variability usually encountered in mass cytometry data from human subjects. The random effect terms induce marker correlations in the posterior distribution. The LKJ prior is an unrestricted prior and thus induces an unrestricted posterior distribution over correlation matrices. Additionally,
the standard deviations can account for overdispersion similar to the negative binomial distribution in the univariate case.

\subsubsection{Posterior Inference}

The parameters of interest are
$$
\boldsymbol{\theta} =
\{
\boldsymbol{\beta},
\boldsymbol{\sigma}^{\text{cond[1]}}, \boldsymbol{\sigma}^{\text{cond}[i]=2}, \boldsymbol{\sigma}^{\text{donor}},
\boldsymbol{\Omega}^{\text{cond}[i]=1}, \boldsymbol{\Omega}^{\text{cond}[i]=2}, \boldsymbol{\Omega}^{\text{donor}}
\},
$$
and we need to sample from the posterior distribution $p(\boldsymbol{\theta}|\boldsymbol{y}) = p(\boldsymbol{y}|\boldsymbol{\theta}) p(\boldsymbol{\theta}) / p(\boldsymbol{y})$ given the data $\boldsymbol{y}$.
The posterior distribution is not analytically tractable. We therefore resort to Markov chain Monte Carlo simulations. We code the model in Stan \citep{carpenter2017stan} and run Hamiltonian Monte Carlo \citep{neal2011mcmc} with 8 chains in parallel. We assess convergence using traceplots; see \R markdown reports in the supplementary material for details. To speed up convergence, we initialize the HMC sampler with maximum likelihood estimates from independent Poisson regressions for the fixed effects, and sample standard deviations and correlation matrices for hyperparameters of the random effects.

\subsubsection{Goodness of Fit}
\label{sec:goodness_of_fit}

\begin{figure}[t]
\centering
\includegraphics[width=\linewidth]{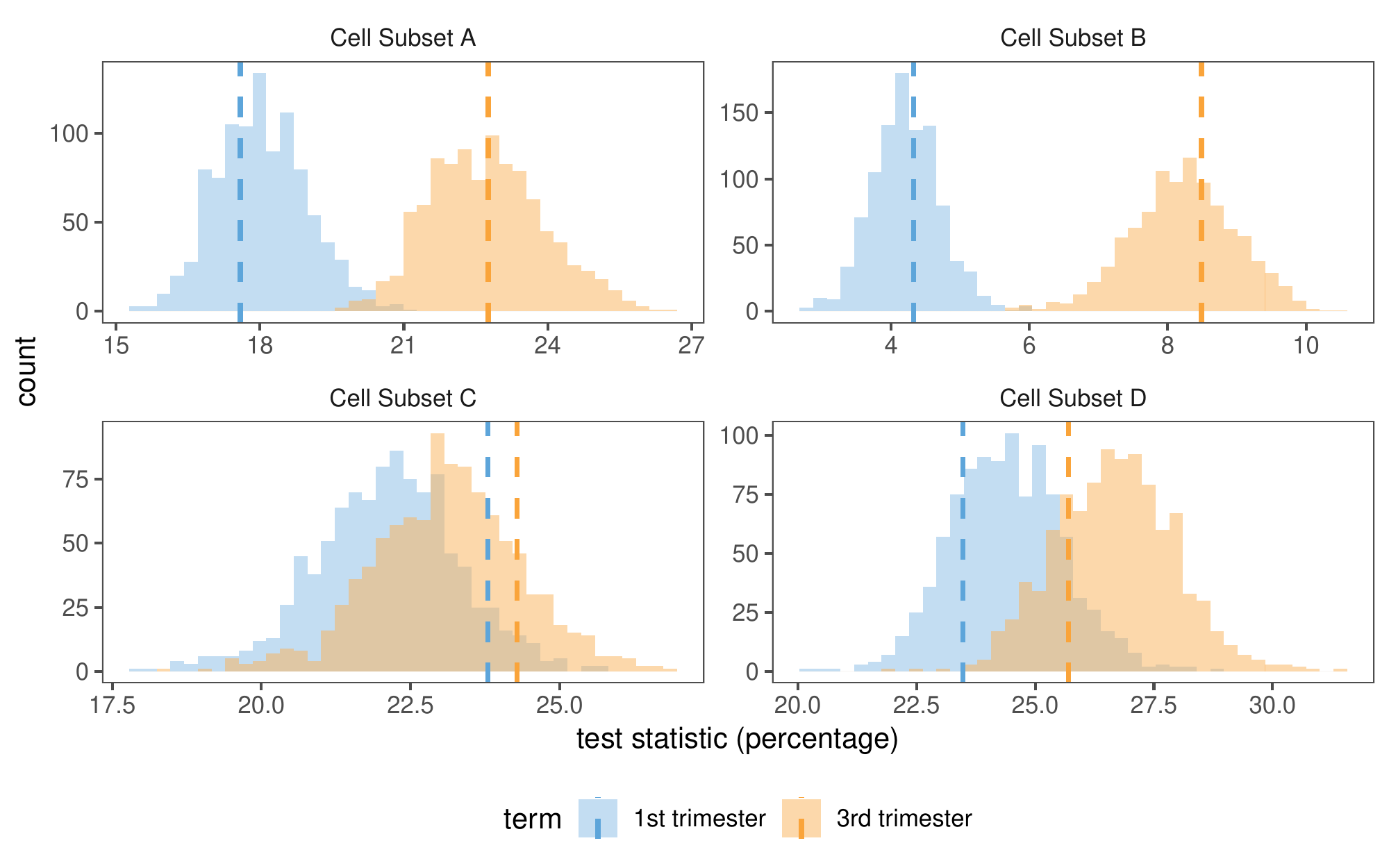}
\caption{Goodness of fit using marginal posterior predictive checks.
\textbf{Cell Subset~A}: Percentage of cells that have larger than median expression of pSTAT1, pSTAT3, and pSTAT5.
\textbf{Cell Subset~B}: Percentage of cells that have larger than median expression of pSTAT1, and lower than median expression of pSTAT3 and pSTAT5.
\textbf{Cell Subset~C}: Percentage of cells that have zero expression of pERK1/2 and larger than median expression of pMAPKAPK2.
\textbf{Cell Subset~D}: Percentage of cells that have nonzero expression of pERK1/2 and larger than median expression of pMAPKAPK2.
The medians are taken over both conditions.}
\label{fig:goodness_of_fit}
\end{figure}
We evaluate the fit of our model using four test statistics that measure model adequacy. For the first test statistic, we calculate the overall median marker expression for pSTAT1, pSTAT3, and pSTAT5. We use the median to define dim (less than the median) and bright (greater than median) cells. We calculate the percentage of bright pSTAT1, pSTAT3, and pSTAT5 cells. For the second test statistic, we calculate the percentage of pSTAT1 bright, pSTAT3 dim, and pSTAT5 dim cells. For the third and fourth test, we calculate the percentage of non-expressed and expressed pERK1/2 and bright pMAPKAPK2 cells, respectively. Figure \ref{fig:goodness_of_fit} shows the comparison of the posterior predictive distribution of this test statistic overlayed with the actual observed test statistic as dashed lines. Observed test statistics in the bulk of the distribution indicate a good model fit.

\subsubsection{Posterior Summary}
\label{sec:posterior_summaries}

We compare pairwise marker correlations and assess if they increase from first to third trimester. We define the posterior probability $p_{ij}$ of larger correlation between marker $i$ and $j$ from $K$ posterior samples as ($I(\cdot)$ is an indicator function)
\[
\hat{p}_{ij} = \frac{1}{K} \sum_{k=1}^K I(\Omega^{\text{cond[i]=2}}_{ijk} > \Omega^{\text{cond}[i]=1}_{ijk}).
\]

\subsection{Logistic Linear Mixed Model}

Now, we change our perspective and predict the experimental condition from protein marker expressions. We use experimental conditions as response variables (transformed; see Section \ref{sub:variance_stabilizing}) and marker expression as explanatory variables. The response variable $y_i$ is a binary variable encoding experimental condition as zero or one. In our reanalysis, $y_i = 0$ and $y_i = 1$ are first and third trimester, respectively. We model the response variable as a Bernoulli random variable with probability $\pi_i$ for each cell. Then we use the $\operatorname{logit}$ link to relate the linear model to binary responses. The linear model predicts the logarithm of the odds of the $i$th cell being from the third instead of the first trimester. The linear model has one coefficient per protein marker $\beta_{i1},\dots,\beta_{iJ}$ (in our reanalysis $J = 10$ proteins markers) and an intercept. If $\pi_i$ is 0.5 then the cell could have come from either first or third trimesters with equal probability. What we observe are the protein marker expressions $\boldsymbol{x}_i$. For each cell we measure $J$ protein markers. If $\pi_i$ is either very close to zero or one, then the cell is strongly representative of a cell observed under the first or third trimester, respectively. The $\pi_i$ values are not observed directly, only $y_i$ and $\boldsymbol{x}_i$ are observed. So, $\pi_i$ has to be estimated from the data. We add flexibility by allowing a mixed effect term in the standard logistic regression model. The covariates $\boldsymbol{x}_i$ are the same as in the fixed effect term, the regression coefficients $\boldsymbol{u}_{\text{donor}[i]}$ vary by donor. Thus allowing markers of each donor and cell type to have varying coefficients. Each cell $i$ maps to a donor indexed by $\text{donor}[i]$. Formally,
\[
\begin{aligned}
y_{i} & \sim \operatorname{Bernoulli}(\pi_{i}) \\
\log\left(\frac{\pi_i}{1-\pi_i}\right) & = \boldsymbol{x}_i^T\boldsymbol{\beta} + \boldsymbol{x}_{i}^T\boldsymbol{u}_{\text{donor}[i]}.
\end{aligned}
\]

We model coefficients using a multivariate normal distribution with covariance matrix decomposed into standard deviations $\boldsymbol{\sigma}^{\text{donor}}$ and correlation matrix $\boldsymbol{\Omega}^{\text{donor}}$. Formally,
\[
\begin{aligned}
\boldsymbol{u}_{\text{donor}[i]} & \sim \operatorname{Normal}\left(\boldsymbol{0}, \operatorname{diag}(\boldsymbol{\sigma}^{\text{donor}}) \, \boldsymbol{\Omega}^{\text{donor}} \, \operatorname{diag}(\boldsymbol{\sigma}^{\text{donor}}) \right).
\end{aligned}
\]
The mixed effect model is a compromise between two extremes. First, we could estimate separate regression coefficients for each donor. This corresponds to random effects modeled as a multivariate normal distribution with infinite standard deviations as there is no constrain on how coefficients are related to each other. Second, we could pool all donors into one group and forget about the donor information. This corresponds to random effects modeled as a multivariate normal distribution with zero standard deviations as there is no additional variability besides the fixed effect term. We note that modeling the additional variability is important, as we noticed that donor-to-donor variability is higher than celltype-to-celltype variability in a wide range of mass cytometry experiments.

We use the same priors as for PLMM. The interpretation of the priors on the regression coefficient $\boldsymbol{\beta}$ are less intuitive than in PLMM. Since the prior is weak, the data will overwhelm the prior. Simulation experiments show little impact of varying the prior on~$\boldsymbol{\beta}$. Formally,
\[
\begin{aligned}
\beta_{ij} & \sim \text{Normal}(0,7^2) \\
\boldsymbol{\sigma}^{\text{donor}} & \sim \text{Half-Cauchy}(0, 2.5) \\
\boldsymbol{\Omega}^{\text{donor}} & \sim \text{LKJ}(1).
\end{aligned}
\]

\begin{figure}[t]
\centering
\includegraphics[width=\linewidth]{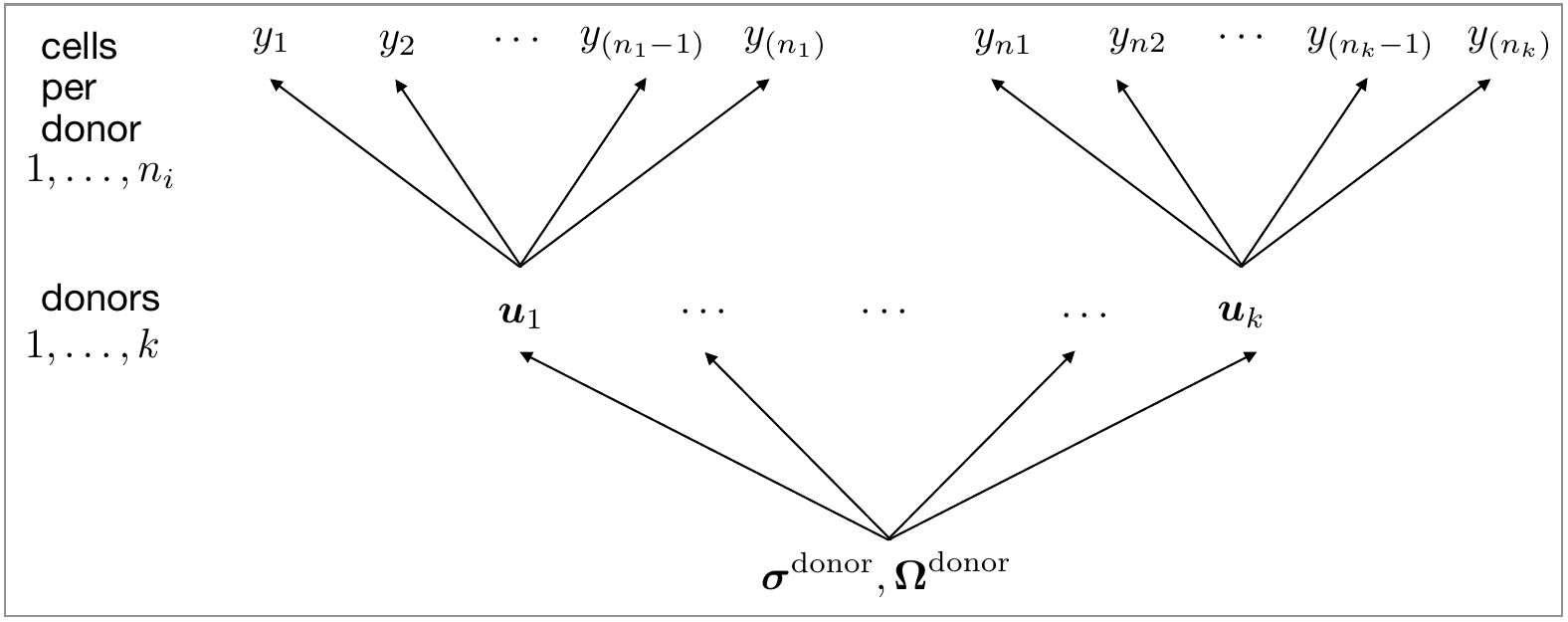}
\caption{A graphical representation of LLMM.}
\label{fig:graphical_llmm}
\end{figure}
Figure \ref{fig:graphical_plmm} displays the LLMM schematically.

\subsubsection{Posterior Inference}

The parameters of interest are $\boldsymbol{\theta} = \{ \boldsymbol{\beta}, \boldsymbol{\sigma}^{\text{donor}}, \boldsymbol{\Omega}^{\text{donor}} \}$,
and we sample from the posterior distribution $p(\boldsymbol{\theta}|\boldsymbol{y})$ given the data $\boldsymbol{y}$ using Hamiltonian Monte Carlo as in PLMM.

We also provide a fast approximation in our \CytoGLMM~package using the method of moments \citep{perry2017fast}. This corresponds to approximate improper flat priors on all parameters. We have successfully used this approach in two recent immunology studies \citep{kronstad2018differential,legars2018cd38}.

\subsubsection{Marker Expression Transformation}
\label{sub:variance_stabilizing}

We assume that marker expressions have been transformed using variance stabilizing transformations to account for heteroskedasticity. We recommend the use of the hyperbolic sine transformation. This transformation assumes a two-component model for the measurement error \citep{rocke1995two,huber2003parameter} where small counts are less noisy than large counts. Intuitively, this corresponds to a noise model with additive and multiplicative noise depending on the magnitude of the marker expression; see \cite{holmes2019modern} for details.

\section{Results}
\label{sec:results}

We follow the gating scheme detailed in \cite{aghaeepour2017immune} and define 12 cell types using the \R package \textsf{openCyto}~\citep{finak2014opencyto}:
memory CD4 positive T cells (CD4+Tmem), naive CD4 positive T cells (CD4+Tnaive),
memory CD8 positive T cells (CD8+Tmem), naive CD8 positive T cells (CD8+Tnaive),
$\gamma\delta$T cells (gdT), regulatory T memory cells (Tregsmem), regulatory T naive cells (Tregsnaive), B cells, classical monocytes (cMC), intermediate monocytes (intMC), non-classical monocytes (ncMC), and Natural Killer cells (NK).

Out of the 32 protein markers measured on each cell, the authors defined 22 markers as gating markers, and 10 as functional markers. The functional markers are pSTAT1, pSTAT3, pSTAT5, pNF$\kappa$B, total I$\kappa$B, pMAPKAPK2, pP38, prpS6, pERK1/2, and pCREB (in plots Greek symbols are replaced by Latin symbols). More details on the gating template is available in the \CytoGLMM~workflow in the supplementary material.

In the interest of expositional clarity, we focus our analysis
on NK cells by filtering the data frame (Figure \ref{fig:data_structure_after_gating}). The same analysis carries over to other cell types.

We run eight parallel Markov chains for 325 iterations and remove the first 200 iterations during the warm up. This gives a total of 1000 draws from the posterior distribution. More details on convergence diagnostics and traceplots are available in our PLMM and LLMM workflows in the supplementary material.

\subsection{Poisson Log-Normal Mixed Model for NK Cells}

We evaluate the fit of our model using four test statistics calculating percentages of four cell subset; see Section~\ref{sec:goodness_of_fit} for details. Goodness of fit plots (Figure~\ref{fig:goodness_of_fit}) show the posterior distributions of these test statistics with dashed lines indicating the observed statistic. The cell subsets A and B are located in the bulk of the distribution indicating an adequate fit. For the cell subset C and D, our model underestimates and overestimates, respectively.

\begin{figure}
\centering
\includegraphics[width=\linewidth]{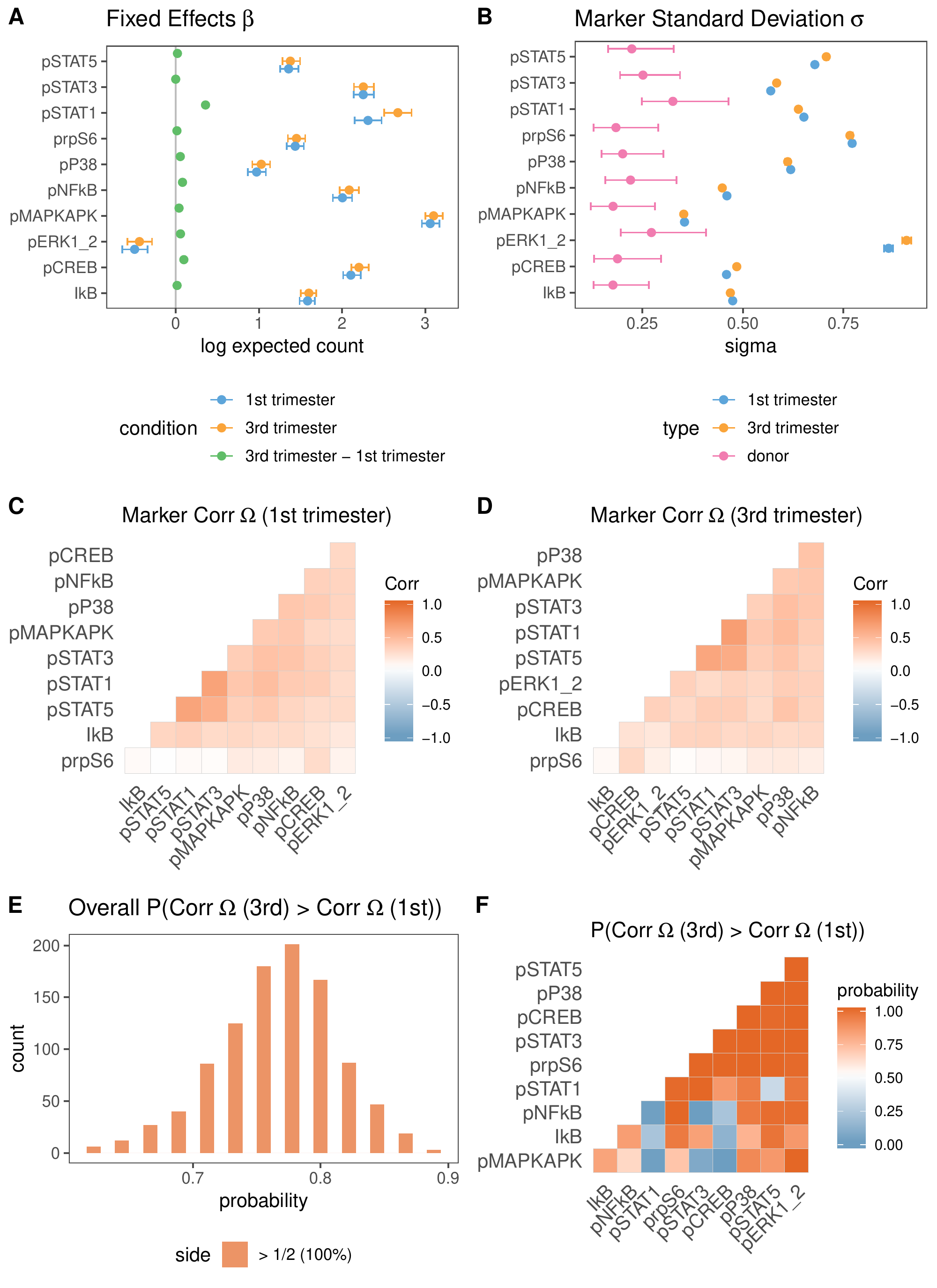}
\caption{\textbf{A}: Marginal posterior median and 95\% posterior credible intervals of coefficient vector for the trimester fixed effects and their differences. \textbf{B}: Standard deviations for the condition random effects. \textbf{C}-\textbf{D}: Marginal posterior median of correlation matrices for the condition random effects. \textbf{E}: Posterior probability of increase in correlations from first to third trimester. \textbf{F}: Pairwise posterior probability of increase in correlations from first to third trimester.}
\label{fig:poisson_results}
\end{figure}
We plot the marginal posterior median and 95\% credible intervals for fixed effects $\boldsymbol{\beta}$ (Figure~\ref{fig:poisson_results}A). In addition to the parameter estimates for the first and third trimesters, we also compute the difference between third and first trimesters. Protein pSTAT1 shows the largest difference between trimesters; an increase of $\exp(0.36) = 1.4$ counts (posterior median). During the first trimester pSTAT1 is expressed between $\exp(2.15) = 8.6$ and $\exp(2.48) = 11.9$ times, and during the third trimester it increases to between $\exp(2.51) = 12.3$ and $\exp(2.84) = 17$ (95\% credible interval). These results account for donor variability. The posterior marker standard deviations $\boldsymbol{\sigma}$ are larger for first and third trimesters than across donors (Figure~\ref{fig:poisson_results}B). Due to the small number of donors (18) relative to the number of cells per donor ($\approx$10,000), the credible intervals are wider in the donor standard deviations. For the posterior marker correlation matrices $\boldsymbol{\Omega}$, we visualize the median marginal posterior correlation (Figure~\ref{fig:poisson_results}C-D). Both during first and third trimester all markers are positively correlated. In particular, pSTAT1, pSTAT3, and pSTAT5 show a strong correlation.
\begin{figure}[t]
\centering
\includegraphics[width=\linewidth]{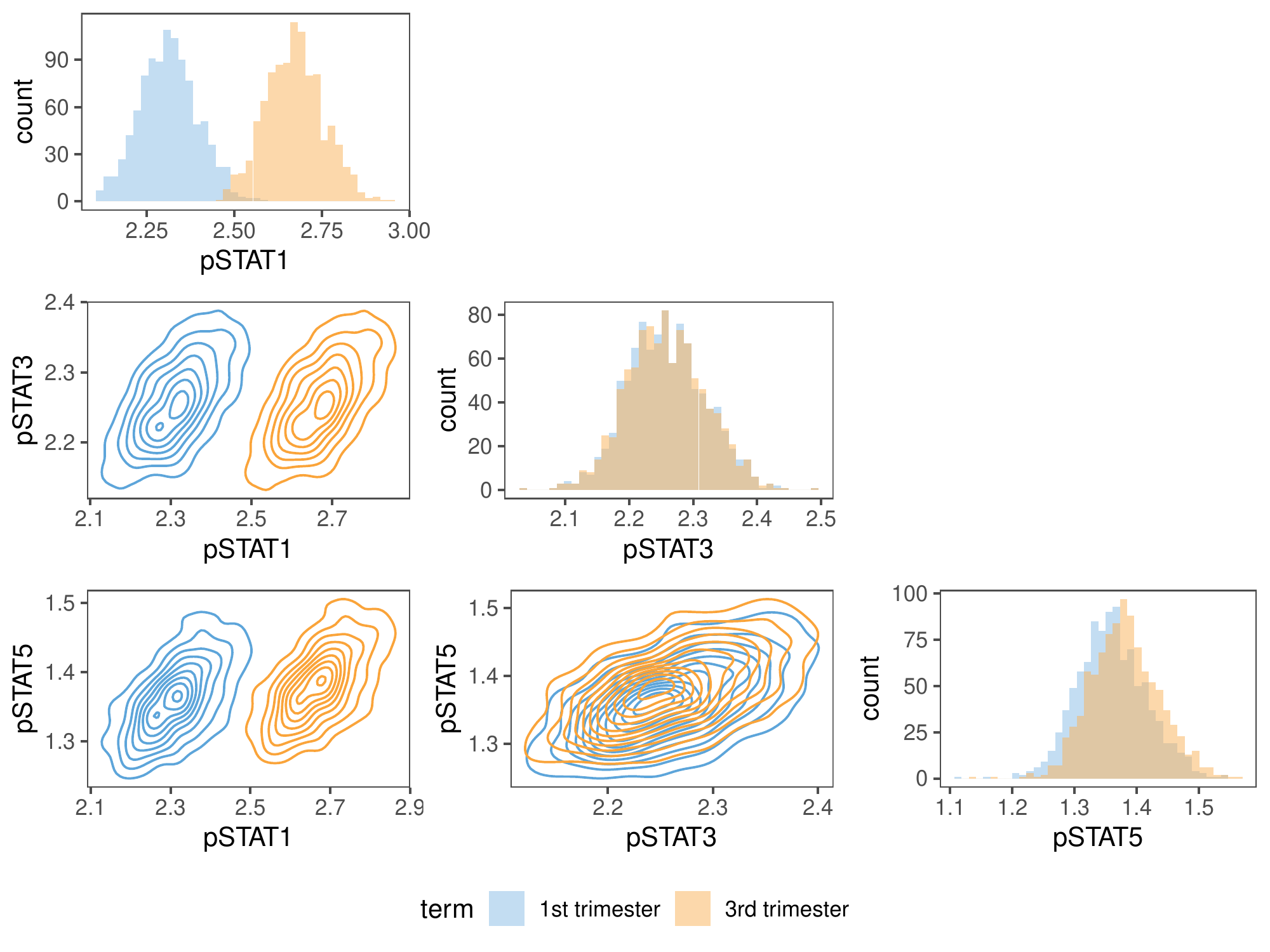}
\caption{Pairs plot of multivariate marginal posterior distribution of three markers.}
\label{fig:poisson_results_multivariate}
\end{figure}
Figure~\ref{fig:poisson_results_multivariate} shows the multivariate marginal joint posterior distribution of pSTAT1, pSTAT3, and pSTAT5.

The condition specific correlations are robust because of the large number of cells. We assess correlation changes from first to third trimester on the global and local scale. On the global scale, we compute the probability
of larger pairwise correlations over all $J(J - 1)/2$ possible correlations. The histogram~(Figure~\ref{fig:poisson_results}E) of these probabilities shows that all pairwise correlations are higher in the third trimester, with a peak around 75-80\%. On the local scale, we compare each pairwise correlation. Figure~\ref{fig:poisson_results}F shows pairwise probabilities of an increase in correlations.

\subsection{Logistic Linear Mixed Model for NK Cells}

\begin{figure}[t]
\centering
\includegraphics[width=\linewidth]{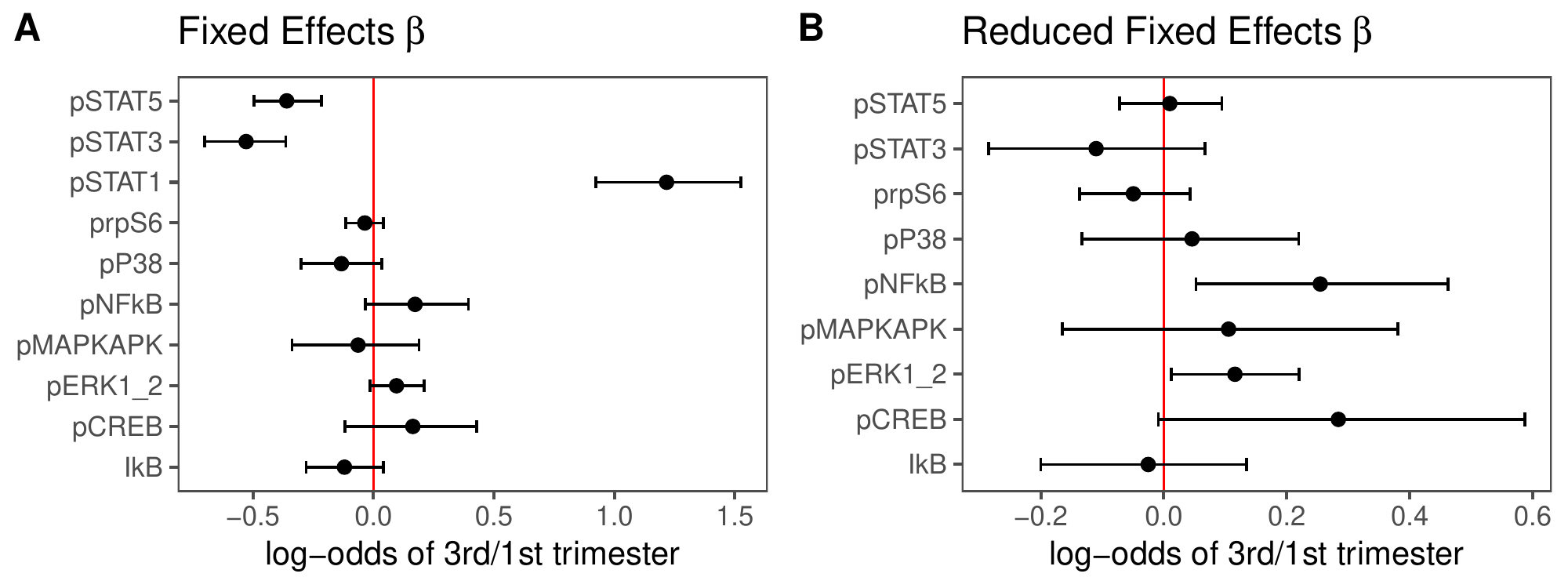}
\caption{Marginal posterior median and 95\% credible intervals for fixed effects. \textbf{A}: All proteins markers. \textbf{B}: Possible collider pSTAT1 removed.}
\label{fig:logistic_regression_results}
\end{figure}
We plot the marginal posterior median and 95\% credible intervals for the fixed effects $\boldsymbol{\beta}$ (Figure~\ref{fig:logistic_regression_results}A). The estimates are on the $\log$-odds scale. We see that pSTAT1 is a strong predictor of the third trimester. This means that a unit increase in the transformed marker expression makes it between $\exp(0.92) = 2.5$ to $\exp(1.52) = 4.6$ (95\% credible interval) more likely to be a cell from the third trimester, while holding the other markers constant. This result coincides with the PLMM result. In contrast, pSTAT3 and pSTAT5 have negative coefficients. This means pSTAT3 and pSTAT5 predict the first trimester, while holding the other markers constant. This result contradicts with the PLMM result. One way to resolve this contradiction is by assuming a collider confounding. The potential collider pSTAT1 would depend on trimester, pSTAT3, and pSTAT5. To test this, we refit the regression model after removing the potential collider pSTAT1. Indeed, the negative coefficients for pSTAT3 and pSTAT5 disappear without pSTAT1 in the model (Figure~\ref{fig:logistic_regression_results}B).

\section{Discussion}
\label{sec:discussion}

In this article, we propose two regression models explaining the relationship between experimental conditions and protein marker expressions. We accounted for donor-to-donor and cell-to-cell variability. We showed that additional modeling of this variability increases the amount of information we can extract from a recent pregnancy study \citep{aghaeepour2017immune}. In analyzing NK cell populations in different subjects during pregnancy, we were able to corroborate an increase of pSTAT1 during the third trimester when samples were stimulated with IFN$\alpha$, as previously reported \citep{aghaeepour2017immune}. We also presented a new way of quantifying
changes in overall and pairwise protein-to-protein correlations during pregnancy, and visualized their multivariate uncertainty. We found that PLMMs and LLMMs can give different results. They agreed on pSTAT1, but reported contradictory results for pSTAT3 and pSTAT5. We resolved these contradictions by assuming a collider confounding. In general, we advocate for using multiple complementary models instead of model selection.

Our two new \R packages \CytoGLMM~and \cytoeffect~are applicable to a wide range of cytometry studies. Besides comparisons on paired samples, where samples are available for the same subject under different stimuli or time points, our PLMM is applicable to unpaired samples, where samples are collected on two separate groups of individuals; e.g.~vaccine studies \citep{sasaki2014distinct}. It is further possible to apply PLMM to more complicated experimental designs by adding more random effect terms; e.g.~twin studies \citep{brodin2015variation}. In contrast, our LLMM is limited to paired samples and simple experimental designs.

Our models provide key advantages over existing approaches. \cite{nowicka2017cytof,weber2018diffcyt,aghaeepour2017immune} compress the data by summarizing by median expression per donor and condition. Such compression hurts the statistical power of the analysis and makes obtaining reliable estimates of marker correlations more difficult. Our models work on the uncompressed full data and models marker correlations explicitly, addressing both aforementioned concerns. \cite{bruggner2014automated,arvaniti2017sensitive} may overfit to the data
because they pool donors prior to the analysis. Our models are more robust to donor-to-donor variability because we explicitly model this variability with random effects. Finally, the PLMM model can incorporate complicated experimental designs, which may be hard for an analysis using hypothesis testing \citep{lun2017testing}. In summary, our reanalysis proves that this family of models avoids loss of information through data compression. In other words, adapting Don Knuth's famous quote about premature optimization in programming: ``premature summarization is the root of all evil (or at least most of it) in statistics.''

The main disadvantages of our complex modeling approach is computation cost. The fact that HMC is inherently a sequential procedure makes it challenging to obtain a speed up without sacrificing accuracy. Our PLMM takes about six days to fit 180,000 cells and 16 donors using eight CPU cores. Future work aims to incorporate recent developments in scaling up HMC \citep{srivastava2018scalable} into our workflow. Another approach is to trade-off the theoretical guarantees of MCMC samplers for time efficiency by using variational inference, as recently introduced by \cite{chiquet2018variational2, chiquet2018variational1} for multivariate Poisson log-normal models. As a check, we conducted the same analysis on a subsample of 1000 cells per donor. We observed that this subsampling step has minimal influence on
fixed effects, marker correlation, and standard deviations, but a larger influence on the marker correlations comparison. We therefore recommend running fast exploratory analysis on smaller subsets of the data as the major source of biological variability is between subjects. For more accurate correlation comparisons, we recommend running the analysis on the full dataset.

Our second limitation is the requirement for gated cell types. To reduce the person-to-person bias of manual gating, we employed the \R package \textsf{openCyto}~\citep{finak2014opencyto}. It is challenging to scale this approach to high dimensional gating schemes. In the future, it will be important to extend our models with a cell type clustering step \citep{silva2017multivariate} and propagate gating uncertainty to the regression model. This could be accomplished by using infinite mixtures from Bayesian nonparametrics to cluster cells and donors into groups of similar immunological profiles. In the univariate setting, \cite{antonelli2016mitigating} showed that more flexible random effect distributions mitigate bias at no or only small prize in efficiency.

Our posterior model checks for PLMM indicated an adequate model fit for most markers except pERK1/2. In the cell subset C, where the pERK1/2's are not expressed, the model predicted a smaller number of cells than in the observed data. The opposite effect occurred in the cell subset D, where the pERK1/2 are nonzero, and the model predicted a larger than observed cell count. In future work, we plan to provide an option to allow for zero-inflated marker counts in the Poisson distribution using mixtures.

We use transformed marker expression as predictors in our LLMM. To propagate uncertainty from this transformation step to the regression step, we plan to add a measurement error model to LLMM.

Donor specific random effects in the pregnancy study are noisy due to the small number of donors. We therefore avoid to interpret the marginal donor specific random effects directly. Instead we consider them nuisance parameters and average over them. We aim to apply our models to studies with larger donor samples size. This will allow us to quantify more precisely the donor-to-donor posterior correlation matrix.

\section*{Supplementary Material: Reproducibility}

Data analysis and results can be completely reproduced following the vignettes on our \R package websites (including installation instructions and source \R markdown files):
\begin{itemize}
\item \url{https://christofseiler.github.io/CytoGLMM}
\item \url{https://christofseiler.github.io/cytoeffect}
\end{itemize}
The data is publicly available and automatically downloaded during the vignettes.

\section*{Acknowledgments}

This work was supported by the National Institutes of Health [U01AI131302 to C.A.B.~and S.H., R56AI124788 to S.H., R21AI130523 to S.H.,
DP1DA046089 to C.A.B., R21AI130532 to C.A.B., R01AI133698 to C.A.B., R21AI135287 to C.A.B., 5T32AI007290-29 to L.M.K., TL1TR001084 to E.V., T32AI007502 to E.V., 1F32AI126674 to L.J.S.]; an A.P. Giannini fellowship [to L.M.K.]; and a Stanford Child Health Research Institute postdoctoral fellowship [to M.L.G.]. C.A.B.~is the Tashia and John Morgridge Endowed Faculty Scholar in Pediatric Translational Medicine from the Maternal Child Health Research Institute, and a Chan Zuckerberg Investigator.

\bibliographystyle{rss}
\bibliography{mixed_models}

\begin{thebibliography}{57}
\expandafter\ifx\csname natexlab\endcsname\relax\def\natexlab#1{#1}\fi
\expandafter\ifx\csname url\endcsname\relax
  \def\url#1{\texttt{#1}}\fi
\expandafter\ifx\csname urlprefix\endcsname\relax\def\urlprefix{URL: }\fi

\bibitem[{Aghaeepour et~al.(2017)Aghaeepour, Ganio, Mcilwain, Tsai, Tingle,
  Van~Gassen, Gaudilliere, Baca, McNeil, Okada, Ghaemi, Furman, Wong, Winn,
  Druzin, El-Sayed, Quaintance, Gibbs, Darmstadt, Shaw, Stevenson, Tibshirani,
  Nolan, Lewis, Angst and Gaudilliere}]{aghaeepour2017immune}
Aghaeepour, N., Ganio, E.~A., Mcilwain, D., Tsai, A.~S., Tingle, M.,
  Van~Gassen, S., Gaudilliere, D.~K., Baca, Q., McNeil, L., Okada, R., Ghaemi,
  M.~S., Furman, D., Wong, R.~J., Winn, V.~D., Druzin, M.~L., El-Sayed, Y.~Y.,
  Quaintance, C., Gibbs, R., Darmstadt, G.~L., Shaw, G.~M., Stevenson, D.~K.,
  Tibshirani, R., Nolan, G.~P., Lewis, D.~B., Angst, M.~S. and Gaudilliere, B.
  (2017) An immune clock of human pregnancy.
\newblock \textit{Science Immunology}, \textbf{2}, eaan2946.

\bibitem[{Aghaeepour et~al.(2011)Aghaeepour, Nikolic, Hoos and
  Brinkman}]{aghaeepour2011rapid}
Aghaeepour, N., Nikolic, R., Hoos, H.~H. and Brinkman, R.~R. (2011) Rapid cell
  population identification in flow cytometry data.
\newblock \textit{Cytometry Part A}, \textbf{79}, 6--13.

\bibitem[{Aitchison and Ho(1989)}]{aitchison1989multivariate}
Aitchison, J. and Ho, C. (1989) The multivariate {P}oisson-log normal
  distribution.
\newblock \textit{Biometrika}, \textbf{76}, 643--653.

\bibitem[{Antonelli et~al.(2016)Antonelli, Trippa and
  Haneuse}]{antonelli2016mitigating}
Antonelli, J., Trippa, L. and Haneuse, S. (2016) Mitigating bias in generalized
  linear mixed models: {T}he case for {B}ayesian nonparametrics.
\newblock \textit{Statistical Science}, \textbf{31}, 80--95.

\bibitem[{Arvaniti and Claassen(2017)}]{arvaniti2017sensitive}
Arvaniti, E. and Claassen, M. (2017) Sensitive detection of rare
  disease-associated cell subsets via representation learning.
\newblock \textit{Nature Communications}, \textbf{8}, 14825.

\bibitem[{Becher et~al.(2014)Becher, Schlitzer, Chen, Mair, Sumatoh, Teng, Low,
  Ruedl, Riccardi-Castagnoli, Poidinger et~al.}]{becher2014high}
Becher, B., Schlitzer, A., Chen, J., Mair, F., Sumatoh, H.~R., Teng, K. W.~W.,
  Low, D., Ruedl, C., Riccardi-Castagnoli, P., Poidinger, M. et~al. (2014)
  High-dimensional analysis of the murine myeloid cell system.
\newblock \textit{Nature Immunology}, \textbf{15}, 1181.

\bibitem[{Bendall et~al.(2011)Bendall, Simonds, Qiu, El-ad, Krutzik, Finck,
  Bruggner, Melamed, Trejo, Ornatsky et~al.}]{bendall2011single}
Bendall, S.~C., Simonds, E.~F., Qiu, P., El-ad, D.~A., Krutzik, P.~O., Finck,
  R., Bruggner, R.~V., Melamed, R., Trejo, A., Ornatsky, O.~I. et~al. (2011)
  Single-cell mass cytometry of differential immune and drug responses across a
  human hematopoietic continuum.
\newblock \textit{Science}, \textbf{332}, 687--696.

\bibitem[{Brodin et~al.(2015)Brodin, Jojic, Gao, Bhattacharya, Angel, Furman,
  Shen-Orr, Dekker, Swan, Butte et~al.}]{brodin2015variation}
Brodin, P., Jojic, V., Gao, T., Bhattacharya, S., Angel, C. J.~L., Furman, D.,
  Shen-Orr, S., Dekker, C.~L., Swan, G.~E., Butte, A.~J. et~al. (2015)
  Variation in the human immune system is largely driven by non-heritable
  influences.
\newblock \textit{Cell}, \textbf{160}, 37--47.

\bibitem[{Bruggner et~al.(2014)Bruggner, Bodenmiller, Dill, Tibshirani and
  Nolan}]{bruggner2014automated}
Bruggner, R.~V., Bodenmiller, B., Dill, D.~L., Tibshirani, R.~J. and Nolan,
  G.~P. (2014) Automated identification of stratifying signatures in cellular
  subpopulations.
\newblock \textit{Proceedings of the National Academy of Sciences},
  \textbf{111}, E2770--E2777.

\bibitem[{Carpenter et~al.(2017)Carpenter, Gelman, Hoffman, Lee, Goodrich,
  Betancourt, Brubaker, Guo, Li and Riddell}]{carpenter2017stan}
Carpenter, B., Gelman, A., Hoffman, M.~D., Lee, D., Goodrich, B., Betancourt,
  M., Brubaker, M., Guo, J., Li, P. and Riddell, A. (2017) Stan: {A}
  probabilistic programming language.
\newblock \textit{Journal of Statistical Software}, \textbf{76}.

\bibitem[{Chan et~al.(1988)Chan, Ng and Hui}]{chan1988simple}
Chan, J., Ng, C. and Hui, P. (1988) A simple guide to the terminology and
  application of leucocyte monoclonal antibodies.
\newblock \textit{Histopathology}, \textbf{12}, 461--480.

\bibitem[{Chen et~al.(2016)Chen, Lau, Wong, Newell, Poidinger and
  Chen}]{chen2016cytofkit}
Chen, H., Lau, M.~C., Wong, M.~T., Newell, E.~W., Poidinger, M. and Chen, J.
  (2016) Cytofkit: {A} {B}ioconductor package for an integrated mass cytometry
  data analysis pipeline.
\newblock \textit{PLoS Computational Biology}, \textbf{12}, e1005112.

\bibitem[{Chib and Winkelmann(2001)}]{chib2001markov}
Chib, S. and Winkelmann, R. (2001) Markov {C}hain {M}onte {C}arlo analysis of
  correlated count data.
\newblock \textit{Journal of Business \& Economic Statistics}, \textbf{19},
  428--435.

\bibitem[{Chiquet et~al.(2018{\natexlab{a}})Chiquet, Mariadassou and
  Robin}]{chiquet2018variational2}
Chiquet, J., Mariadassou, M. and Robin, S. (2018{\natexlab{a}}) Variational
  inference for sparse network reconstruction from count data.
\newblock \textit{arXiv preprint arXiv:1806.03120}.

\bibitem[{Chiquet et~al.(2018{\natexlab{b}})Chiquet, Mariadassou, Robin
  et~al.}]{chiquet2018variational1}
Chiquet, J., Mariadassou, M., Robin, S. et~al. (2018{\natexlab{b}}) Variational
  inference for probabilistic {P}oisson {PCA}.
\newblock \textit{The Annals of Applied Statistics}, \textbf{12}, 2674--2698.

\bibitem[{Finak et~al.(2009)Finak, Bashashati, Brinkman and
  Gottardo}]{finak2009merging}
Finak, G., Bashashati, A., Brinkman, R. and Gottardo, R. (2009) Merging mixture
  components for cell population identification in flow cytometry.
\newblock \textit{Advances in Bioinformatics}, \textbf{2009}.

\bibitem[{Finak et~al.(2014)Finak, Frelinger, Jiang, Newell, Ramey, Davis,
  Kalams, De~Rosa and Gottardo}]{finak2014opencyto}
Finak, G., Frelinger, J., Jiang, W., Newell, E.~W., Ramey, J., Davis, M.~M.,
  Kalams, S.~A., De~Rosa, S.~C. and Gottardo, R. (2014) Opencyto: {A}n open
  source infrastructure for scalable, robust, reproducible, and automated,
  end-to-end flow cytometry data analysis.
\newblock \textit{PLoS Computational Biology}, \textbf{10}, e1003806.

\bibitem[{Finck et~al.(2013)Finck, Simonds, Jager, Krishnaswamy, Sachs, Fantl,
  Pe'er, Nolan and Bendall}]{finck2013normalization}
Finck, R., Simonds, E.~F., Jager, A., Krishnaswamy, S., Sachs, K., Fantl, W.,
  Pe'er, D., Nolan, G.~P. and Bendall, S.~C. (2013) Normalization of mass
  cytometry data with bead standards.
\newblock \textit{Cytometry Part A}, \textbf{83}, 483--494.

\bibitem[{Ge and Sealfon(2012)}]{ge2012flowpeaks}
Ge, Y. and Sealfon, S.~C. (2012) {flowPeaks}: {A} fast unsupervised clustering
  for flow cytometry data via {K}-means and density peak finding.
\newblock \textit{Bioinformatics}, \textbf{28}, 2052--2058.

\bibitem[{Gelman et~al.(2006)}]{gelman2006prior}
Gelman, A. et~al. (2006) Prior distributions for variance parameters in
  hierarchical models (comment on article by {B}rowne and {D}raper).
\newblock \textit{Bayesian Analysis}, \textbf{1}, 515--534.

\bibitem[{Hern{\'a}n and Robins(2019)}]{hernan2019causal}
Hern{\'a}n, M. and Robins, J. (2019) \textit{Causal Inference}.
\newblock Boca Raton: Chapman \& Hall/CRC, forthcoming.

\bibitem[{Holmes and Huber(2019)}]{holmes2019modern}
Holmes, S. and Huber, W. (2019) \textit{Modern Statistics for Modern Biology}.
\newblock Cambridge University Press.

\bibitem[{Huber et~al.(2003)Huber, von Heydebreck, S{\"u}ltmann, Poustka and
  Vingron}]{huber2003parameter}
Huber, W., von Heydebreck, A., S{\"u}ltmann, H., Poustka, A. and Vingron, M.
  (2003) Parameter estimation for the calibration and variance stabilization of
  microarray data.
\newblock \textit{Statistical Applications in Genetics and Molecular Biology},
  \textbf{2}.

\bibitem[{Imbens and Rubin(2015)}]{imbens2015causal}
Imbens, G.~W. and Rubin, D.~B. (2015) \textit{Causal Inference in Statistics,
  Social, and Biomedical Sciences}.
\newblock Cambridge University Press.

\bibitem[{Kronstad et~al.(2018)Kronstad, Seiler, Vergara, Holmes and
  Blish}]{kronstad2018differential}
Kronstad, L.~M., Seiler, C., Vergara, R., Holmes, S.~P. and Blish, C.~A. (2018)
  Differential induction of {IFN}-$\alpha$ and modulation of {CD112} and {CD54}
  expression govern the magnitude of {NK} cell {IFN}-$\gamma$ response to
  influenza {A} viruses.
\newblock \textit{The Journal of Immunology}, \textbf{201}, 2117--2131.

\bibitem[{Le~Gars et~al.(2018)Le~Gars, Seiler, Kay, Bayless, Starosvetsky,
  Moore, Shen-Orr, Aziz, Dekker, Khatri et~al.}]{legars2018cd38}
Le~Gars, M., Seiler, C., Kay, A., Bayless, N., Starosvetsky, E., Moore, L.,
  Shen-Orr, S., Aziz, N., Dekker, C., Khatri, P. et~al. (2018) {CD38} is a key
  regulator of enhanced {NK} cell immune responses during pregnancy through its
  role in immune synapse formation.
\newblock \textit{bioRxiv}, 349084.

\bibitem[{Levine et~al.(2015)Levine, Simonds, Bendall, Davis, El-ad, Tadmor,
  Litvin, Fienberg, Jager, Zunder et~al.}]{levine2015data}
Levine, J.~H., Simonds, E.~F., Bendall, S.~C., Davis, K.~L., El-ad, D.~A.,
  Tadmor, M.~D., Litvin, O., Fienberg, H.~G., Jager, A., Zunder, E.~R. et~al.
  (2015) Data-driven phenotypic dissection of {AML} reveals progenitor-like
  cells that correlate with prognosis.
\newblock \textit{Cell}, \textbf{162}, 184--197.

\bibitem[{Lewandowski et~al.(2009)Lewandowski, Kurowicka and
  Joe}]{lewandowski2009generating}
Lewandowski, D., Kurowicka, D. and Joe, H. (2009) Generating random correlation
  matrices based on vines and extended onion method.
\newblock \textit{Journal of Multivariate Analysis}, \textbf{100}, 1989--2001.

\bibitem[{Lo et~al.(2009)Lo, Hahne, Brinkman and Gottardo}]{lo2009flowclust}
Lo, K., Hahne, F., Brinkman, R.~R. and Gottardo, R. (2009) flowclust: {A}
  bioconductor package for automated gating of flow cytometry data.
\newblock \textit{BMC Bioinformatics}, \textbf{10}, 145.

\bibitem[{Lun et~al.(2017)Lun, Richard and Marioni}]{lun2017testing}
Lun, A.~T., Richard, A.~C. and Marioni, J.~C. (2017) Testing for differential
  abundance in mass cytometry data.
\newblock \textit{Nature Methods}, \textbf{14}, 707.

\bibitem[{McElreath(2019)}]{mcelreath2016statistical}
McElreath, R. (2019) \textit{Statistical Rethinking$^2$}.
\newblock publisher unknown, draft of second edn.

\bibitem[{Meehan et~al.(2014)Meehan, Walther, Moore, Orlova, Meehan, Parks,
  Ghosn, Philips, Mitsunaga, Waters et~al.}]{meehan2014autogate}
Meehan, S., Walther, G., Moore, W., Orlova, D., Meehan, C., Parks, D., Ghosn,
  E., Philips, M., Mitsunaga, E., Waters, J. et~al. (2014) Autogate: automating
  analysis of flow cytometry data.
\newblock \textit{Immunologic Research}, \textbf{58}, 218--223.

\bibitem[{Naim et~al.(2014)Naim, Datta, Rebhahn, Cavenaugh, Mosmann and
  Sharma}]{naim2014swift}
Naim, I., Datta, S., Rebhahn, J., Cavenaugh, J.~S., Mosmann, T.~R. and Sharma,
  G. (2014) {SWIFT} -- {S}calable clustering for automated identification of
  rare cell populations in large, high-dimensional flow cytometry datasets,
  {P}art 1: Algorithm design.
\newblock \textit{Cytometry Part A}, \textbf{85}, 408--421.

\bibitem[{Neal et~al.(2011)}]{neal2011mcmc}
Neal, R.~M. et~al. (2011) {MCMC} using {H}amiltonian dynamics.
\newblock \textit{Handbook of Markov Chain Monte Carlo}, \textbf{2}, 2.

\bibitem[{Nowicka et~al.(2017)Nowicka, Krieg, Weber, Hartmann, Guglietta,
  Becher, Levesque and Robinson}]{nowicka2017cytof}
Nowicka, M., Krieg, C., Weber, L., Hartmann, F., Guglietta, S., Becher, B.,
  Levesque, M. and Robinson, M. (2017) Cytof workflow: {D}ifferential discovery
  in high-throughput high-dimensional cytometry datasets [version 2; referees:
  2 approved].
\newblock \textit{F1000Research}, \textbf{6}.

\bibitem[{Orlova et~al.(2018)Orlova, Herzenberg and
  Walther}]{orlova2018science}
Orlova, D.~Y., Herzenberg, L.~A. and Walther, G. (2018) Science not art:
  statistically sound methods for identifying subsets in multi-dimensional flow
  and mass cytometry data sets.
\newblock \textit{Nature Reviews Immunology}, \textbf{18}, 77.

\bibitem[{Pearl(2009)}]{pearl2009causality}
Pearl, J. (2009) \textit{Causality}.
\newblock Cambridge University Press.

\bibitem[{Perry(2017)}]{perry2017fast}
Perry, P.~O. (2017) Fast moment-based estimation for hierarchical models.
\newblock \textit{Journal of the Royal Statistical Society: Series B
  (Statistical Methodology)}, \textbf{79}, 267--291.

\bibitem[{Peters et~al.(2017)Peters, Janzing and
  Sch{\"o}lkopf}]{peters2017elements}
Peters, J., Janzing, D. and Sch{\"o}lkopf, B. (2017) \textit{Elements of Causal
  Inference: Foundations and Learning Algorithms}.
\newblock MIT Press.

\bibitem[{Qian et~al.(2010)Qian, Wei, Eun-Hyung~Lee, Campbell, Halliley, Lee,
  Cai, Kong, Sadat, Thomson et~al.}]{qian2010elucidation}
Qian, Y., Wei, C., Eun-Hyung~Lee, F., Campbell, J., Halliley, J., Lee, J.~A.,
  Cai, J., Kong, Y.~M., Sadat, E., Thomson, E. et~al. (2010) Elucidation of
  seventeen human peripheral blood {B}-cell subsets and quantification of the
  tetanus response using a density-based method for the automated
  identification of cell populations in multidimensional flow cytometry data.
\newblock \textit{Cytometry Part B: Clinical Cytometry}, \textbf{78}, S69--S82.

\bibitem[{Qiu et~al.(2011)Qiu, Simonds, Bendall, Gibbs~Jr, Bruggner, Linderman,
  Sachs, Nolan and Plevritis}]{qiu2011extracting}
Qiu, P., Simonds, E.~F., Bendall, S.~C., Gibbs~Jr, K.~D., Bruggner, R.~V.,
  Linderman, M.~D., Sachs, K., Nolan, G.~P. and Plevritis, S.~K. (2011)
  Extracting a cellular hierarchy from high-dimensional cytometry data with
  {SPADE}.
\newblock \textit{Nature Biotechnology}, \textbf{29}, 886.

\bibitem[{Rocke and Lorenzato(1995)}]{rocke1995two}
Rocke, D.~M. and Lorenzato, S. (1995) A two-component model for measurement
  error in analytical chemistry.
\newblock \textit{Technometrics}, \textbf{37}, 176--184.

\bibitem[{Rothman et~al.(2008)Rothman, Greenland and Lash}]{rothman2008modern}
Rothman, K.~J., Greenland, S. and Lash, T.~L. (2008) \textit{Modern
  Epidemiology}.
\newblock Lippincott Williams and Wilkins.

\bibitem[{Saeys et~al.(2016)Saeys, Van~Gassen and
  Lambrecht}]{saeys2016computational}
Saeys, Y., Van~Gassen, S. and Lambrecht, B.~N. (2016) Computational flow
  cytometry: {H}elping to make sense of high-dimensional immunology data.
\newblock \textit{Nature Reviews Immunology}, \textbf{16}, 449.

\bibitem[{Samusik et~al.(2016)Samusik, Good, Spitzer, Davis and
  Nolan}]{samusik2016automated}
Samusik, N., Good, Z., Spitzer, M.~H., Davis, K.~L. and Nolan, G.~P. (2016)
  Automated mapping of phenotype space with single-cell data.
\newblock \textit{Nature Methods}, \textbf{13}, 493.

\bibitem[{Sasaki et~al.(2014)Sasaki, Holmes, Albrecht, Garc{\'\i}a-Sastre,
  Dekker, He and Greenberg}]{sasaki2014distinct}
Sasaki, S., Holmes, T.~H., Albrecht, R.~A., Garc{\'\i}a-Sastre, A., Dekker,
  C.~L., He, X.-S. and Greenberg, H.~B. (2014) Distinct cross-reactive {B}-cell
  responses to live attenuated and inactivated influenza vaccines.
\newblock \textit{The Journal of infectious diseases}, \textbf{210}, 865--874.

\bibitem[{Seiler and Holmes(2017)}]{seiler2017multivariate}
Seiler, C. and Holmes, S. (2017) Multivariate heteroscedasticity models for
  functional brain connectivity.
\newblock \textit{Frontiers in Neuroscience}, \textbf{11}, 696.

\bibitem[{Shekhar et~al.(2014)Shekhar, Brodin, Davis and
  Chakraborty}]{shekhar2014automatic}
Shekhar, K., Brodin, P., Davis, M.~M. and Chakraborty, A.~K. (2014) Automatic
  classification of cellular expression by nonlinear stochastic embedding
  {(ACCENSE)}.
\newblock \textit{Proceedings of the National Academy of Sciences},
  \textbf{111}, 202--207.

\bibitem[{Silva et~al.(2017)Silva, Rothstein, McNicholas and
  Subedi}]{silva2017multivariate}
Silva, A., Rothstein, S.~J., McNicholas, P.~D. and Subedi, S. (2017) A
  multivariate {P}oisson-log normal mixture model for clustering transcriptome
  sequencing data.
\newblock \textit{arXiv preprint arXiv:1711.11190}.

\bibitem[{S{\"o}rensen et~al.(2015)S{\"o}rensen, Baumgart, Durek, Gr{\"u}tzkau
  and H{\"a}upl}]{sorensen2015immunoclust}
S{\"o}rensen, T., Baumgart, S., Durek, P., Gr{\"u}tzkau, A. and H{\"a}upl, T.
  (2015) immunoclust -- {A}n automated analysis pipeline for the identification
  of immunophenotypic signatures in high-dimensional cytometric datasets.
\newblock \textit{Cytometry Part A}, \textbf{87}, 603--615.

\bibitem[{Spirtes et~al.(2000)Spirtes, Glymour, Scheines, Heckerman, Meek,
  Cooper and Richardson}]{spirtes2000causation}
Spirtes, P., Glymour, C.~N., Scheines, R., Heckerman, D., Meek, C., Cooper, G.
  and Richardson, T. (2000) \textit{Causation, Prediction, and Search}.
\newblock MIT Press.

\bibitem[{Spitzer et~al.(2015)Spitzer, Gherardini, Fragiadakis, Bhattacharya,
  Yuan, Hotson, Finck, Carmi, Zunder, Fantl et~al.}]{spitzer2015interactive}
Spitzer, M.~H., Gherardini, P.~F., Fragiadakis, G.~K., Bhattacharya, N., Yuan,
  R.~T., Hotson, A.~N., Finck, R., Carmi, Y., Zunder, E.~R., Fantl, W.~J.
  et~al. (2015) An interactive reference framework for modeling a dynamic
  immune system.
\newblock \textit{Science}, \textbf{349}, 1259425.

\bibitem[{Srivastava et~al.(2018)Srivastava, Li and
  Dunson}]{srivastava2018scalable}
Srivastava, S., Li, C. and Dunson, D.~B. (2018) Scalable {B}ayes via barycenter
  in wasserstein space.
\newblock \textit{The Journal of Machine Learning Research}, \textbf{19},
  312--346.

\bibitem[{Van~Gassen et~al.(2015)Van~Gassen, Callebaut, Van~Helden, Lambrecht,
  Demeester, Dhaene and Saeys}]{van2015flowsom}
Van~Gassen, S., Callebaut, B., Van~Helden, M.~J., Lambrecht, B.~N., Demeester,
  P., Dhaene, T. and Saeys, Y. (2015) {FlowSOM}: {U}sing self-organizing maps
  for visualization and interpretation of cytometry data.
\newblock \textit{Cytometry Part A}, \textbf{87}, 636--645.

\bibitem[{Weber et~al.(2018)Weber, Nowicka, Soneson and
  Robinson}]{weber2018diffcyt}
Weber, L.~M., Nowicka, M., Soneson, C. and Robinson, M.~D. (2018) diffcyt:
  {D}ifferential discovery in high-dimensional cytometry via high-resolution
  clustering.
\newblock \textit{bioRxiv}.
\newblock
  \urlprefix\url{https://www.biorxiv.org/content/early/2018/06/18/349738}.

\bibitem[{Weber and Robinson(2016)}]{weber2016comparison}
Weber, L.~M. and Robinson, M.~D. (2016) Comparison of clustering methods for
  high-dimensional single-cell flow and mass cytometry data.
\newblock \textit{Cytometry Part A}, \textbf{89}, 1084--1096.

\bibitem[{Zare et~al.(2010)Zare, Shooshtari, Gupta and Brinkman}]{zare2010data}
Zare, H., Shooshtari, P., Gupta, A. and Brinkman, R.~R. (2010) Data reduction
  for spectral clustering to analyze high throughput flow cytometry data.
\newblock \textit{BMC Bioinformatics}, \textbf{11}, 403.

\end{thebibliography}
\end{document}